\documentclass[conference]{IEEEtran}
\IEEEoverridecommandlockouts
\usepackage{cite}
\usepackage{amsmath,amssymb,amsfonts}
\usepackage{algorithmic}
\usepackage{graphicx}
\usepackage{textcomp}
\usepackage{xcolor}
\def\BibTeX{{\rm B\kern-.05em{\sc i\kern-.025em b}\kern-.08em
    T\kern-.1667em\lower.7ex\hbox{E}\kern-.125emX}}
\begin{document}

\title{Towards a Dynamic Composability Approach for using Heterogeneous Systems in Remote Sensing\\
\thanks{Identify applicable funding agency here. If none, delete this.}
}

\author{\IEEEauthorblockN{Ilkay Altintas}
\IEEEauthorblockA{\textit{University of California, San Diego}\\
La Jolla, CA, USA \\
ialtintas@ucsd.edu}
\and
\IEEEauthorblockN{Ismael Perez}
\IEEEauthorblockA{\textit{University of California, San Diego}\\
La Jolla, CA, USA \\
i3perez@sdsc.edu}
\and
\IEEEauthorblockN{Dmitry Mishin}
\IEEEauthorblockA{\textit{University of California, San Diego}\\
La Jolla, CA, USA \\
dmishin@sdsc.edu}
\and
\IEEEauthorblockN{Adrien Trouillaud}
\IEEEauthorblockA{\textit{Admiralty}\\
Seattle, WA, USA \\
adrien@admiralty.io}
\and
\IEEEauthorblockN{Christopher Irving}
\IEEEauthorblockA{\textit{University of California, San Diego}\\
La Jolla, CA, USA \\
cirving@sdsc.edu}
\and
\IEEEauthorblockN{John Graham}
\IEEEauthorblockA{\textit{University of California, San Diego}\\
La Jolla, CA, USA \\
jjgraham@ucsd.edu}
\and
\IEEEauthorblockN{Mahidhar Tatineni}
\IEEEauthorblockA{\textit{University of California, San Diego}\\
La Jolla, CA, USA \\
mahidhar@sdsc.edu}
\and
\IEEEauthorblockN{Thomas DeFanti}
\IEEEauthorblockA{\textit{University of California, San Diego}\\
La Jolla, CA, USA \\
tdefanti@ucsd.edu}
\and
\IEEEauthorblockN{Shawn Strande}
\IEEEauthorblockA{\textit{University of California, San Diego}\\
La Jolla, CA, USA \\
strande@sdsc.edu}
\and
\IEEEauthorblockN{Larry Smarr}
\IEEEauthorblockA{\textit{University of California, San Diego}\\
La Jolla, CA, USA \\
lsmarr@ucsd.edu}
\and
\IEEEauthorblockN{Michael L. Norman}
\IEEEauthorblockA{\textit{University of California, San Diego}\\
La Jolla, CA, USA \\
mlnorman@ucsd.edu}
}

\maketitle

\begin{abstract}
Influenced by the advances in data and computing, the scientific practice increasingly involves machine learning and artificial intelligence driven methods which requires specialized capabilities at the system-, science- and service-level in addition to the conventional large-capacity supercomputing approaches. The latest distributed architectures built around the composability of data-centric applications led to the emergence of a new ecosystem for container coordination and integration. However, there is still a divide between the application development pipelines of existing supercomputing environments, and these new dynamic environments that disaggregate fluid resource pools through accessible, portable and re-programmable interfaces. New approaches for dynamic composability of heterogeneous systems are needed to further advance the data-driven scientific practice for the purpose of more efficient computing and usable tools for specific scientific domains. In this paper, we present a novel approach for using composable systems in the intersection between scientific computing, artificial intelligence (AI), and remote sensing domain. We describe the architecture of a first working example of a composable infrastructure that federates Expanse, an NSF-funded supercomputer, with Nautilus, a Kubernetes-based GPU geo-distributed cluster. We also summarize a case study in wildfire modeling, that demonstrates the application of this new infrastructure in scientific workflows: a composed system that bridges the insights from edge sensing, AI and computing capabilities with a physics-driven simulation. 
\end{abstract}

\begin{IEEEkeywords}
 Composable Systems, Kubernetes, Multi-Cluster Federation, Artificial Intelligence, Remote Sensing
\end{IEEEkeywords}

\section{Introduction}
The last two decades have seen a tremendous increase in the volume and complexity of data originating from a variety of sources, including large scientific instruments, simulations, social media, and the Internet of Things (IoT). The need to analyze such data has driven equally impressive gains in deep learning techniques and computer architecture. As a consequence, researchers have developed scientific workflows that run on a continuum of resources, providing near real-time capabilities for processing these data streams. 

In addition, availability of new processing units and computing environments enables unprecedented improvements in scientific software libraries and services optimized for specialized processors. While some processors can be more amenable to simulations in certain fields of study (e.g., 1000s of processing units and high-memory machines required to scale the recursive algorithms of cosmology and physics), other types of processing units, e.g., graphics-processing units (GPUs), Field Programmable Gate Arrays (FPGAs) and tensor processing units (TPUs), can accelerate the data-intensive matrix-multiplication algorithms used in machine learning (ML) or neural networks for computer vision. However, traditional high-performance computing infrastructure is mostly built around enabling computational simulations with limited capabilities for supporting the ever-growing needs of data-driven science, which requires convergence of such heterogeneous capabilities. 

Such a convergence between big data, artificial intelligence (AI) and computational advances has the potential for unprecedented advances in the scientific endeavor \cite{osti_1604756}, pushing the boundaries of existing systems and requiring seamless and permanent integration of data source instruments and the computing ecosystem to enable closed-loop workflows that facilitate linkages between observation, experimentation and simulation based approaches. 

\begin{figure}
    \centering
    \includegraphics[width=0.47\textwidth]{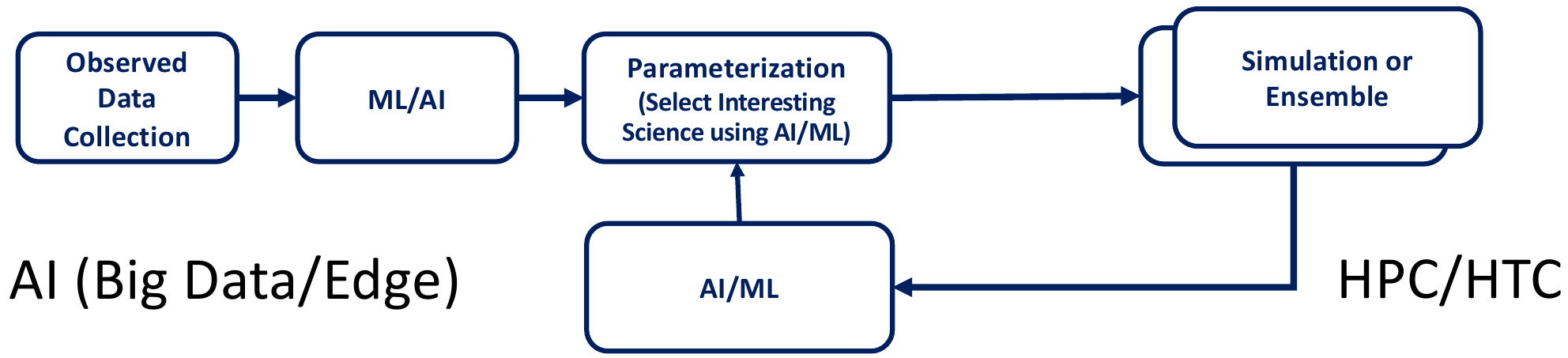}
    \vspace{-2ex}
    \caption{An AI-integrated heterogeneous workflow.}
    \label{fig:hwf}
\end{figure}

As an example of closing the loop between observation and simulation, Figure~\ref{fig:hwf} illustrates a typical dynamic data-driven application workflow. Such applications requires analyzing real-time data to identify parameterization of a downstream ensemble of models or simulations, the results of which can further be analyzed for a closed-loop re-parameterization, e.g., akin to the data-driven wildfire modeling workflows in WIFIRE \cite{wifire15}. When we look at the computing ecosystem required by such a workflow, the first half of the workflow requires AI capabilities involving big data and edge computing instead of a conventional high-performance system. For example, a number of ML methods can run using regular GPUs and other processing units like tensor and intelligence units, and edge devices might require down-scaling of existing methods. Solving such problems require the ability to use and balance extensible heterogeneous capabilities and high performance capacity.

The workflows with high-heterogeneity needs as the one described above can benefit from seamless and dynamic composition of diverse and distributed compute, storage and networking resources as well as integration of heterogeneous software services. We refer to this distributed system integration, which goes beyond the bounds of a single computing system, as a composable system. The simplest definition of a \textit{composable system} is “\textit{an infrastructure that composes into an optimal configuration to meet application demands}”, leveraging an ecosystem of middleware components and microservices to enable efficient management, coordination and allocation of physical resources. The strength of a composable system is its ability to allow application workflows to leverage specialized hardware or other components without requiring that they be physically deployed within a single supercomputer or cluster. 

A conventional system to solve a scientific problem would consists of some hierarchy of components melded together in a pipeline which can be executed on discrete data and system components. In contrast, a composable system is a form in which a conventional system has fulfilled a series of higher-order requirements such that the system can now be considered "composed", and becomes a computational resource basin through which data, users and solutions flow. 
These requirements are summarized as:
\begin{itemize}

\item \textbf{Interoperability:} The use of containers, permanently deployed microservices, and continuous integration/delivery capabilities enables a constant hosting and evolution of data-intensive workloads that can be composed within applications composed of many steps \cite{Bajcsy}.  
\item \textbf{Dynamic Scalability:} The integrated application can be composed of modules and services which scale independently from one another, executing on distributed platforms that best fit their system requirements.   
Microservices can be deployed and updated ad infinitum, allowing for constant real-time updates of the context of each module.
\item \textbf{Interactive Access:} Applications are accessible by users through high-level programming environments, e.g., Python and Spark. Resources for interacting with software and interfaces such as JupyterLabs are always available even if the underlying resource pools dynamically change over time.  
\item \textbf{Performance Measurement:} Continuous measurements related to service execution performance and system availability is needed for dynamic scheduling and re-scheduling throughout the workflow execution cycle.
\end{itemize}
\vspace{-1ex}
Composable systems responding to these requirements unlock the potential and capabilities to provide scalable solutions, and to enable extensible solutions involving the rapid processor advances and growing broader needs of data-driven science. 

Container orchestration frameworks \cite{7036275} like Kubernetes \cite{k8} enable on demand scale-out of disparate containerized application stacks on optimized and distributed computing and storage environments interconnected through high-speed low-latency networks. In such a software defined infrastructure, the hardware and storage is dynamically configured based on varying demand.

\textbf{Contributions. }In this paper, we present a novel approach for using composable systems in the scientific computing domain. We describe the architecture of a first working example of a composable infrastructure that federates Expanse, an NSF-funded supercomputer, with Nautilus, a Kubernetes-based GPU geo-distributed cluster which is a part of a bigger federation of clusters TNRP (Towards National Research Platform). We also summarize a case study in wildfire modeling that demonstrates the application of this new infrastructure in a scientific workflow: a composed system that bridges the insights from edge sensing and computing capabilities with a physics-driven simulation.

\textbf{Outline. }The rest of this paper is organized as follows. In Section 2, we introduce a federated composable architecture using Kubernetes. Section 3 describes the remote sensing case study built on top of the composable hardware and storage architecture described in Section 2. We review related work in Section 4 and conclude in Section 5.

\section{A Composable Architecture using Microservices}

Multi-node distributed software architectures built out of many containerized independent components communicating through APIs is called a \textit{microservice architecture}. For service-oriented applications, such architectures have several benefits including reliability and horizontal scalability compared to monolithic architectures. With its ability to deploy and scale containerized microservice instances as \textit{Kubernetes Pods}, Kubernetes provides capabilities to specify system requirements (e.g., for memory and CPU) and enables applications integration using several microservice instances running across a cluster of nodes, referred to as a \textit{Kubernetes Cluster}. Kubernetes provides a great flexibility in the types of hardware that can be mixed in the same cluster. Assigning labels to the nodes and reporting the available hardware on the node allows users to conveniently choose the resources they would like to use. 

In this section, we first describe two independent Kubernetes clusters, namely Nautilus and Enthalphy, which are respectively constructed on top of the distributed Pacific Research Platform (PRP) \cite{prp} and the Expanse computer \cite{expanse} at the San Diego Supercomputer Center. We then describe how these two Kubernetes clusters were federated into a composable infrastructure for integrated execution of heterogeneous workflow applications. 

\subsection{Nautilus: A Distributed Big Data Processor}

The rapid rise of containers and Kubernetes technology in the last years and wide adoption of those by both the private sector and scientific community allowed the PRP's prototype of Kubernetes cluster built on top of network-measuring nodes to become a widely adopted computing resource provider for a broad scientific community and the frontier for testing new technologies like FPGAs, new GPUs, other types of accelerators, and new software projects. 

The Nautilus hyper-converged cluster is the focus of the PRP's facilities and support for measurement and monitoring as well as distributed computing and shared storage. Currently, the Nautilus cluster aggregates a wide range and several generations of GPUs from older NVIDIA 1080's to newest NVIDIA A100, several FPGA boards, wide range of nodes from tiny 4-core to 96-core 4TB RAM and more, connected with 10-100GB Science DMZ \cite{dmz} networks and allowing all this hardware to work together and use several kinds of geo-distributed persistent storage virtualized by Kubernetes storage drivers. To provide application users adequate storage for their datasets, the PRP includes an extensible storage cloud, currently at 2.3 PB of 10-18-TB disk drives embedded as data capacitors using Ceph \cite{ceph}. 

\begin{figure*}
    \centering
    \includegraphics[width=0.66\textwidth]{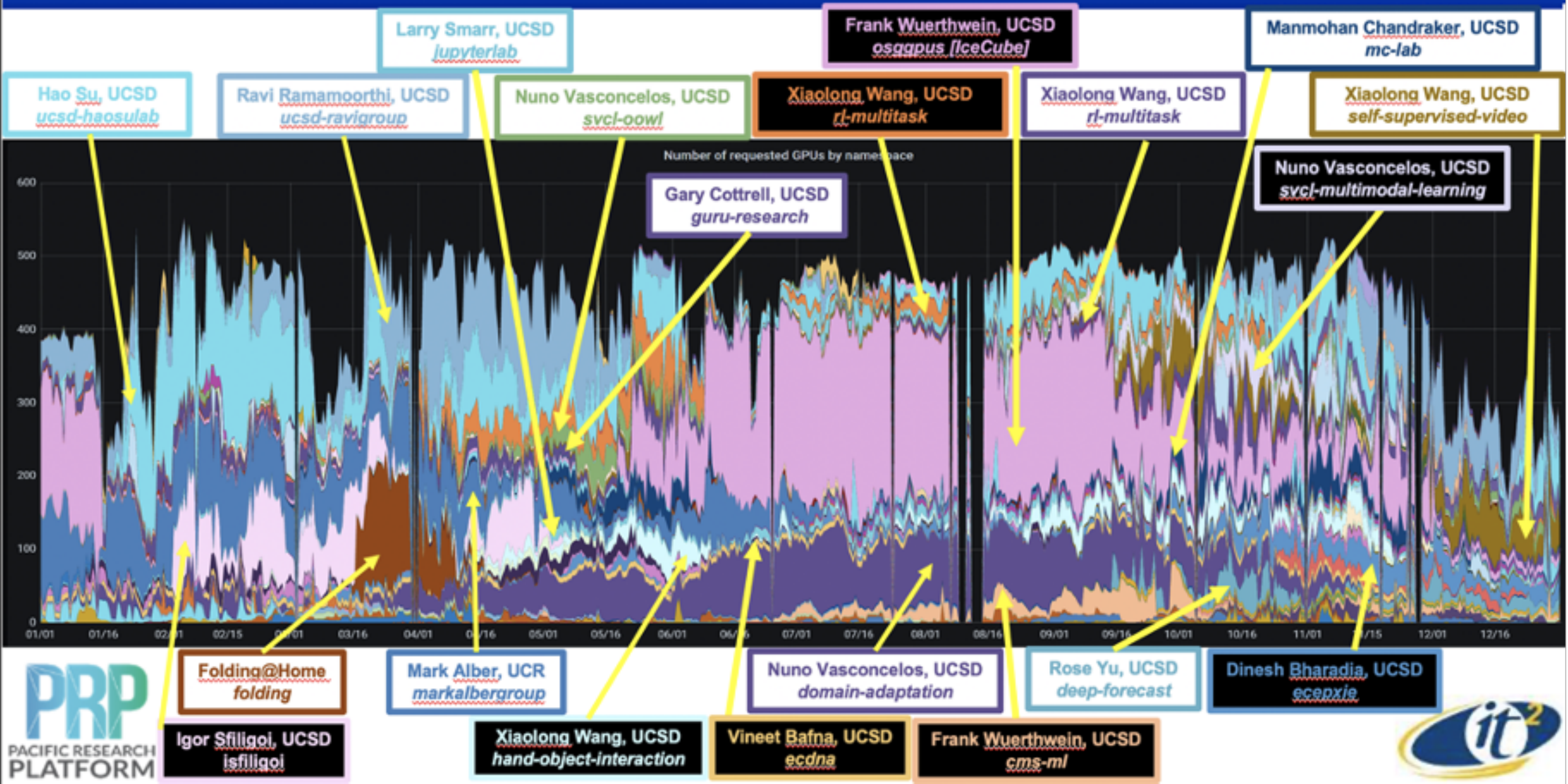}
    \caption{Nautilus GPU usage by group (users from different Kubernetes namespaces) in 2020 on a stacked line graph. GPU usage is shown from January to December (x-axis) and counting the number of GPUs used (y-axis) by groups. The yellow arrow annotations with color-coded show the different groups and their usage.}
    \label{fig:nu}
\end{figure*}

Ceph storage platform implements object storage on a distributed computer cluster, and provides interfaces for object-, block- and file-level storage. Ceph aims primarily for completely distributed operation without a single point of failure, and is scalable to the exabyte level. Ceph replicates data and makes it fault-tolerant, using commodity hardware and requiring no specific hardware support. As a result of its design, the system is ultimately both self-healing and self-managing, minimizing administration time and other costs. 

\begin{figure*}
    \centering
    \includegraphics[width=0.76\textwidth]{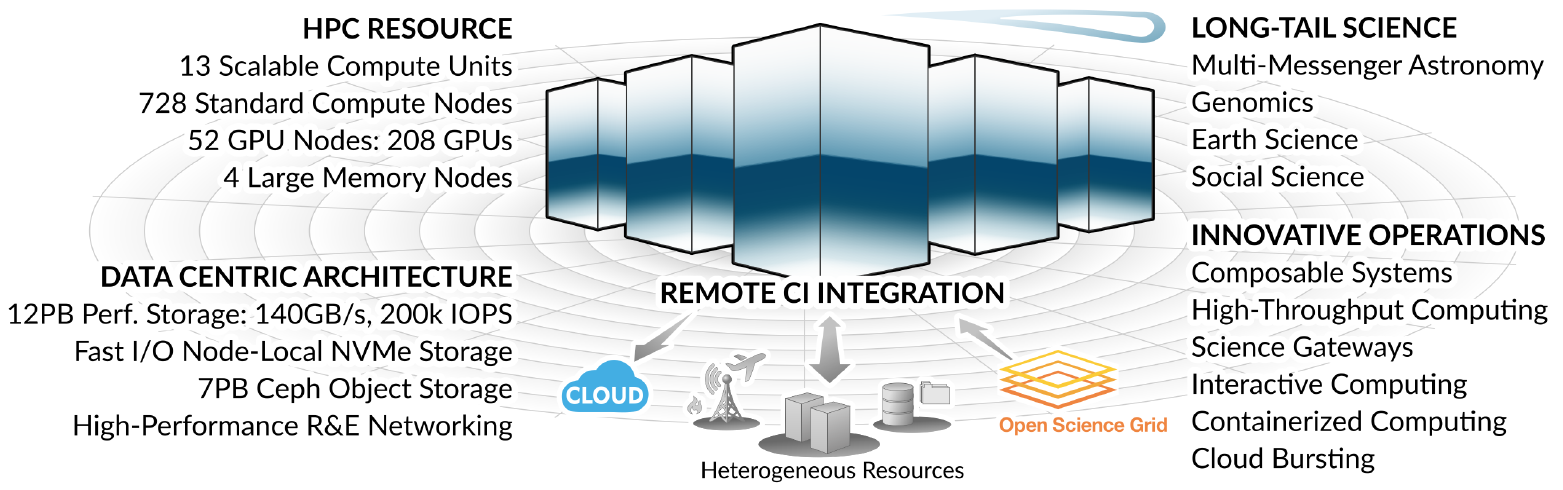}
    \caption{Overview of the capabilities of Expanse. With the strong focus to target the long-tail of science, Expanse will provide a new feature to compose with Kubernetes clusters via a federation layer and the main focus of this paper.}
    \label{fig:exp-ov}
\end{figure*}

\begin{figure}
    \centering
    \includegraphics[width=0.47\textwidth]{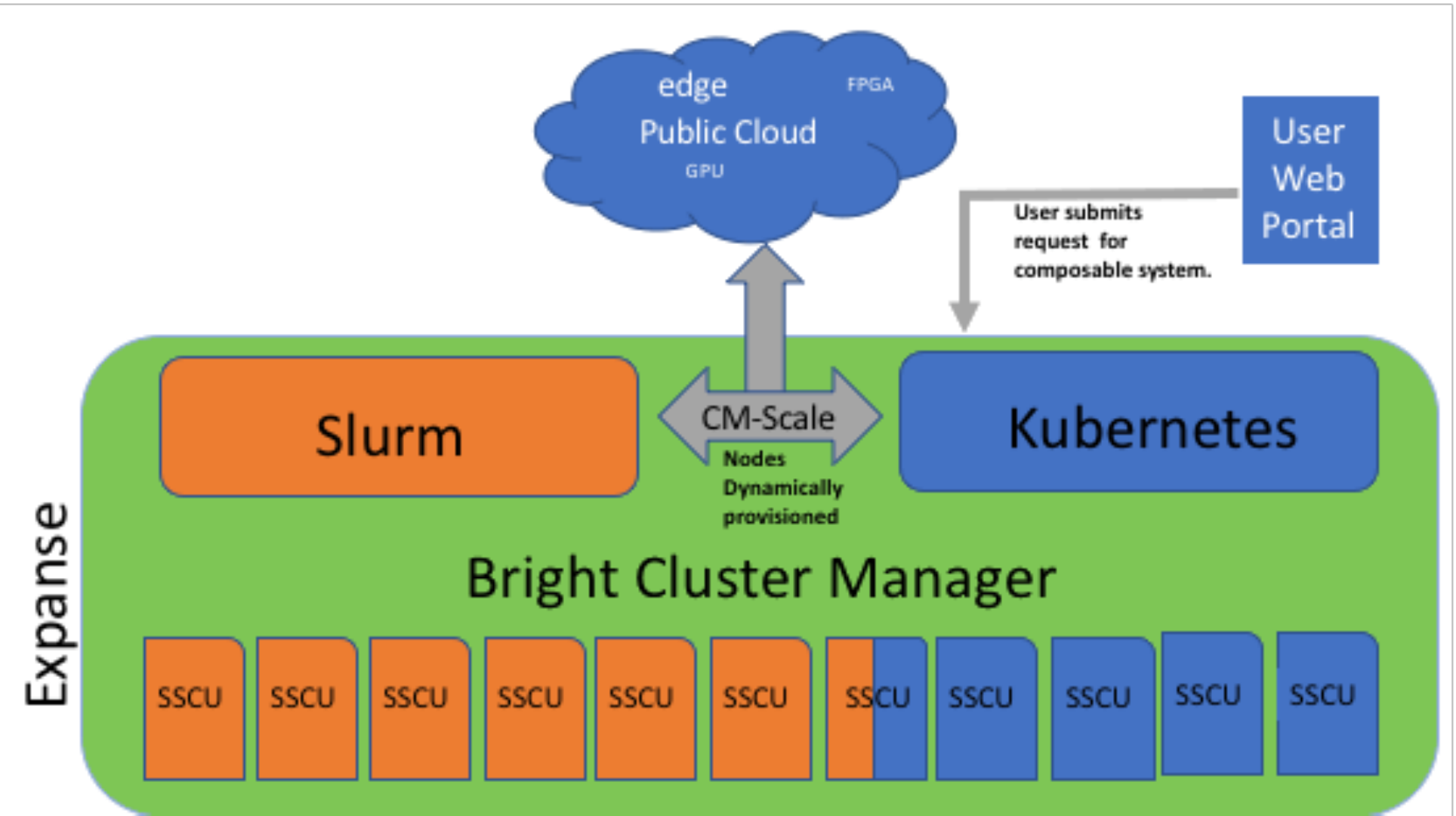}
    \caption{The Composable Systems Framework of Expanse. The cluster is dynamically scaled into the slurm (shown in Orange) and Kubernetes (shown in blue) partitions via cm-scale. }
    \label{fig:exp}
\end{figure}

In addition, to support the needs of application users to carry out ML on their large datasets, Nautilus maintains machine-learning FIONA8s (low-cost, flash memory-based data servers with 8 single-precision/32-bit GPUs) \cite{fiona}, making the PRP a low-cost ML network for big data analysis, as well as a data sharing network \cite{snacs2019}. This enables researchers to deploy, share and jointly control their own computational and data resources. Figure~\ref{fig:nu} shows the number of represented GPUs per namespace (i.e., virtual Kubernetes clusters) representing applications ranging from neural networks, time-series analysis to protein folding. 

However, even with such flexibility, the system is pushing the boundaries of cross-architecture connectivity, e.g., adding certain kinds of hardware to the cluster is not easy and can be close to impossible. Some nodes for IoT devices are too tiny to even handle the standard monitoring services installed on all nodes in the cluster. Some nodes (e.g., ARM architectures for reduced instruction set computing) are more easily managed as a separate cluster. Some clusters just do not fit in Nautilus and have to be managed by other teams, like Expanse and several others. For these reasons, federation is the primary way of expansion and composition for Nautilus. At the same time, the wide range of Nautilus hardware can be made available to other clusters through federation, requiring a composition of different clusters to enable integrated heterogeneous applications with nodes available in Nautilus.

\subsection{Expanse: Computing Without Boundaries}

In the Fall of 2020, the San Diego Supercomputer Center launched its newest National Science Foundation (NSF)-funded supercomputer, Expanse \cite{expanse}. Expanse supports thousands of users of batch-oriented and science gateway computing. As summarized in Figure~\ref{fig:exp-ov}, Expanse also provides new advanced capabilities and innovative operations that enables research increasingly dependent upon heterogeneous and distributed resources composed into integrated and highly usable cyberinfrastructure.

Each of Expanse's 728 standard compute nodes are powered by two 64-core AMD EPYC 7742 processors and contain 256 GB of DDR4 memory, while each of the 52 GPU node contains four NVIDIA V100s (32 GB/GPU), connected via NVLINK, and dual 20-core Intel Xeon 6248 CPUs. Expanse also has four 2 TB large memory nodes.
 
The entire system, integrated by Dell, is organized into 13 SDSC Scalable Compute Units (SSCUs), comprising 56 standard nodes and four GPU nodes, and connected with 100 GB/s HDR InfiniBand. Every Expanse node has access to a 12 PB Lustre parallel file system (provided by Aeon Computing) and 7 PB Ceph Object Store system.

The Expanse cluster is managed using the Bright Computing HPC Cluster management system \cite{bright}, and uses the SLURM workload manager for job scheduling. While the system is suited for modest-scale jobs as few as tens of cores to several hundred cores, Expanse also handles high-throughput computing jobs via integration with the Open Science Grid \cite{osg}, which can have tens of thousands of single-core jobs, and provides connectivity to commercial clouds via the job queuing system. A low-latency interconnect based on Mellanox High Data Rate (HDR) InfiniBand supports a fabric topology optimized for jobs of one to a few thousand cores that require medium-scale parallelism.

One of the key innovations of Expanse is its ability to support composable systems, allowing researchers to create a virtual cluster of resources, for a specific project and then re-compose it as needed. Composable systems workloads that integrate Expanse use Kubernetes through the Bright Cluster Manager \cite{bright}, as depicted in Figure~\ref{fig:exp}. Bright Cluster Manager includes capabilities for configuring, managing and deploying Slurm and Kubernetes-based clusters from a single interface and includes features to dynamically re-provision nodes from the same hardware pool to run either under Slurm or Kubernetes based on resource needs. We name this Kubernetes cluster in Expanse \textit{Enthalpy}.

Bright Cluster Manager automates many of the steps that would otherwise require manual administrative intervention using Kubernetes commands. The cm-scale tool within Bright Cluster Manager is a top-level meta-scheduler that can monitor workloads and dynamically re-purpose a set of compute nodes for Kubernetes. The task of provisioning a new cluster, or potentially joining an existing one is simplified, creating additional possibilities for users to automate their workflows and leverage the existing containerized software repositories.  

\begin{figure*}
    \centering
    \includegraphics[width=0.7\textwidth]{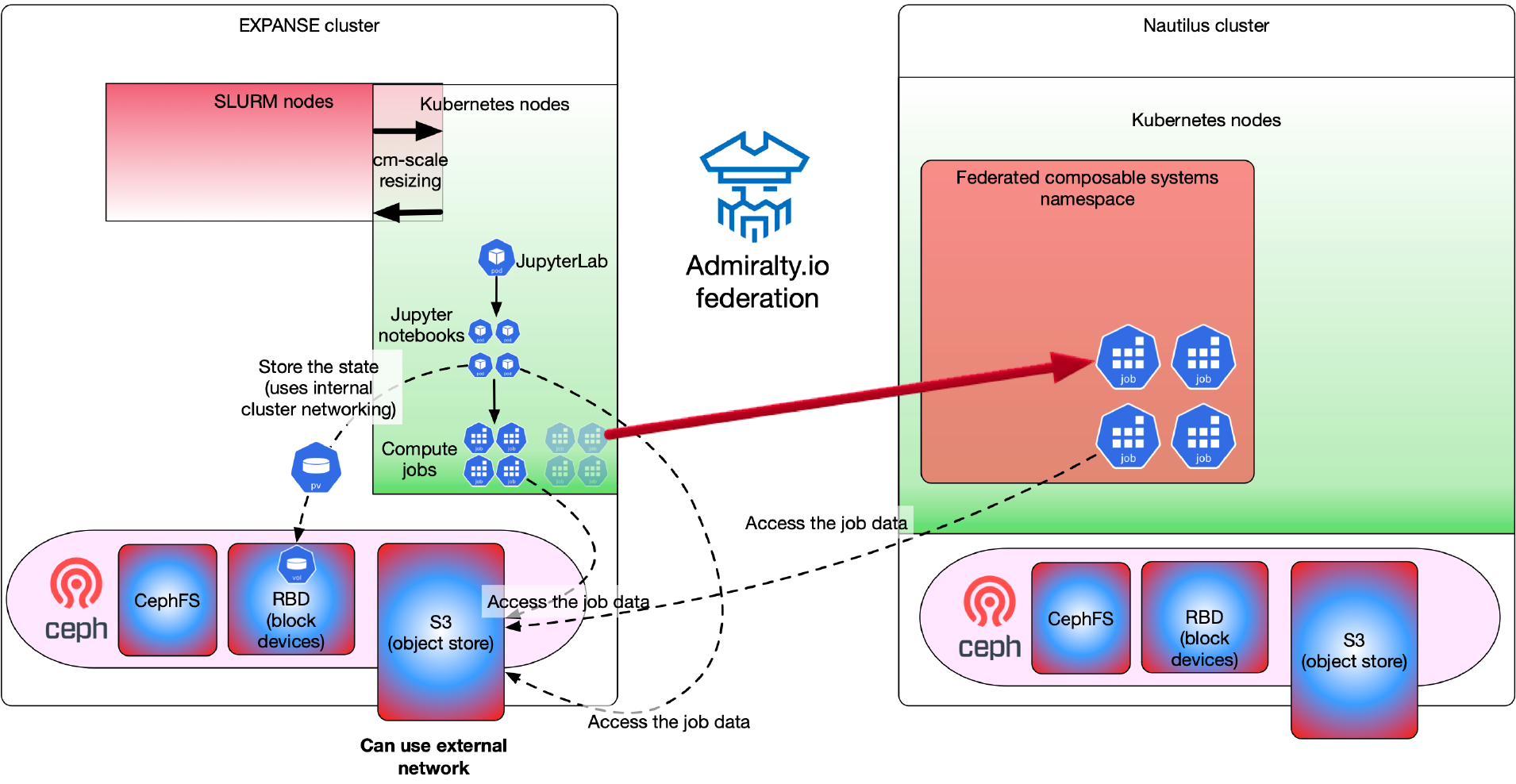}
    \caption{Federation of Enthalpy and Nautilus as a Composable Cluster. Left: The resizable node and storage architecture of the Enthalpy Kubernetes Cluster on Expanse. Right: Nautilus Cluster with the federated composable namespace.}
    \label{fig:fed}
\end{figure*}

The left box titled ``EXPANSE Cluster'' in Figure~\ref{fig:fed} further illustrates the details of the cm-scale resizing of nodes and the Ceph based storage cloud integration. A multi-user JupyterLab \cite{jupyterlab} hub instance running as an interactive application development interface enables deployment of applications on top of the Kubernetes microservice architecture.

Although the task of provisioning Kubernetes clusters is already handled by Bright Cluster Manager, joining other clusters either directly or using federation tools has multiple options (e.g., \cite{consul,admiralty,kubefed}) and required future investigation before Admiralty was used for this purpose.

\begin{figure*}
    \centering
    \includegraphics[width=0.7\textwidth]{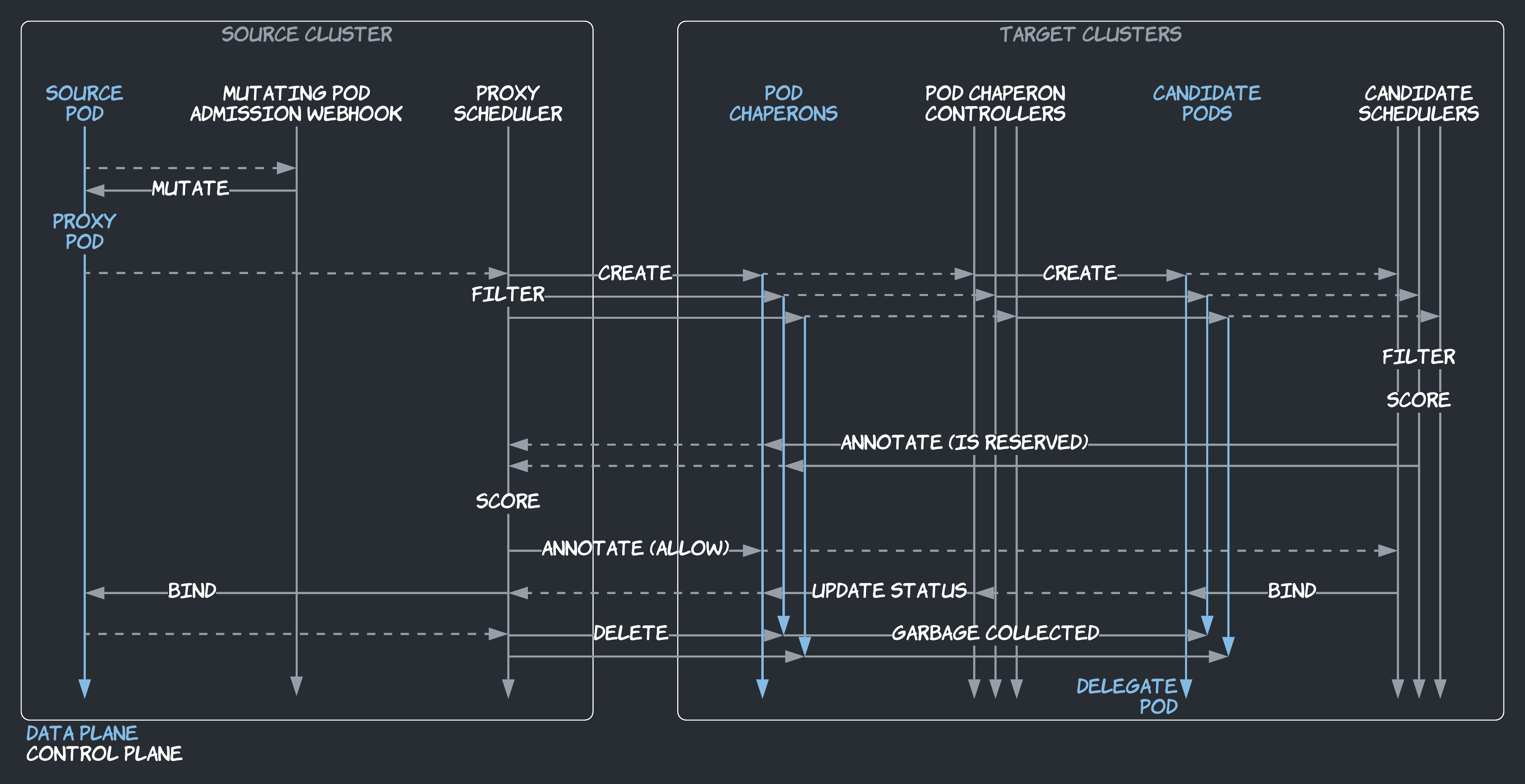}
    \caption{Admiralty scheduling diagram between a source and target clusters with all the events that are happening on both the control and data planes shown in white and blue, respectively.}
    \label{fig:sched}
\end{figure*}

\subsection{Federation Architecture}

Figure~\ref{fig:fed} shows the federation architecture that was used to compose Nautilus and Enthalpy Clusters for seamless workload offloading between these clusters. In this section, we describe this federation architecture using Admiralty \cite{admiralty} and the capabilities offered by Admiralty that made it possible.

Admiralty connects the control planes of Kubernetes clusters. Two clusters are said to be connected as source and target when controllers in the source cluster (e.g., run by the Admiralty agent) communicate with the Kubernetes API of the target cluster. Just like controllers within a single cluster, controllers in either cluster may communicate using the target cluster's Kubernetes API as a message channel. Target clusters are represented by virtual nodes in corresponding source clusters. This is done by using the virtual kubelet - a logical component running inside the cluster which represents a physical node and allows scheduling workloads on a non-existing node. The proxy pods running on a virtual kubelet look like real, but those just clone the state of remote pods and allow watching the state of those from a local cluster.

Pods in the source cluster labeled to be federated are replaced with special proxy pods, which look like regular ones but don't run a workload. Admiralty proxy scheduler connects to remote cluster(s) using credentials obtained during the federation set up. In target cluster a pod chaperon is created. Pod chaperon annotations are used as two-way cross-cluster communication channels between proxy and candidate schedulers to orchestrate scheduling and binding cycles. When scheduling a proxy pod (to bind it to a virtual node), the proxy scheduler doesn't know the hardware specs, policies and current utilization of the target clusters. It does not filter based on aggregate data, which wouldn't be accurate. Instead, it sends candidate pods to all target clusters. The candidate schedulers have all the knowledge required to determine if those pods can be scheduled by using the Scheduling Framework\cite{schedfw} - the special Kubernetes extension making it easy to extend the standard Kubernetes scheduler with custom logic. After scoring the virtual nodes that passed the filter (based on aggregate but good enough data this time), the proxy scheduler elects one candidate pod as the delegate pod. Eventually, the delegate pod is bound, the proxy pod is bound, and all other candidate pods are deleted.

This scheme allows scheduling the pods according to all cluster policies set up by remote cluster admins, which would be impossible if federation layer tried to mimic the actual cluster scheduler. Attempts to do this would always lag behind the actual cluster logic development and it would take too much effort to keep up with the cluster development.

\begin{figure}
    \centering
    \includegraphics[width=0.3\textwidth]{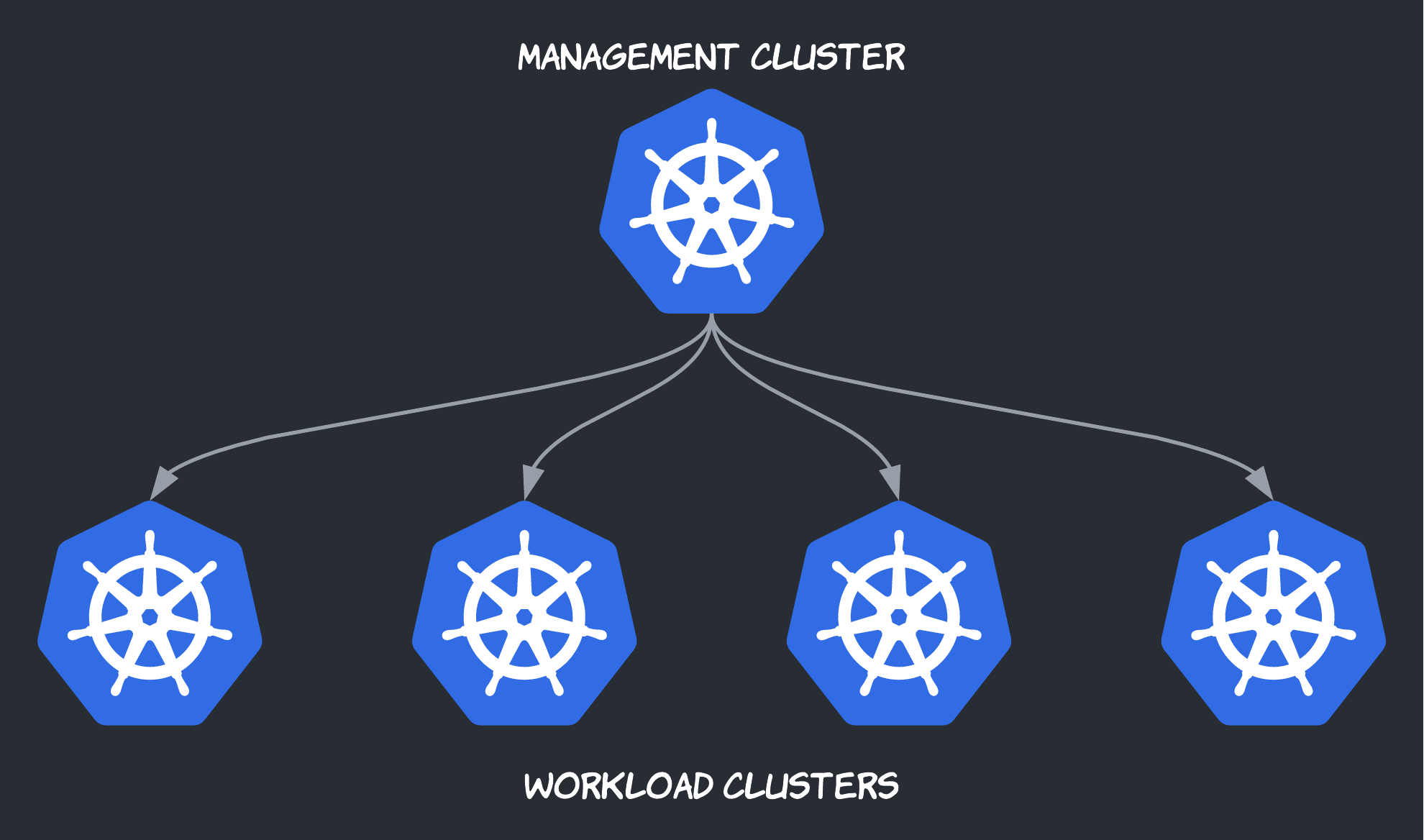}
    \caption{Admiralty federation central control plane}
    \label{fig:central_plane}
\end{figure}

\begin{figure}
    \centering
    \includegraphics[width=0.35\textwidth]{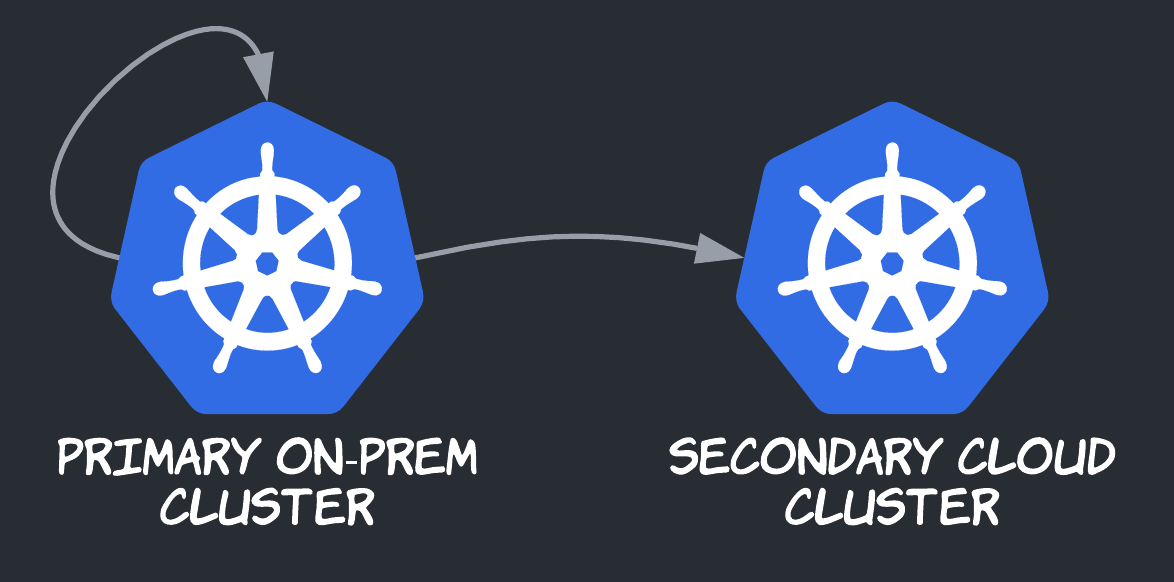}
    \caption{Admiralty federation cloud bursting}
    \label{fig:cloud_burst}
\end{figure}

\begin{figure}
    \centering
    \includegraphics[width=0.25\textwidth]{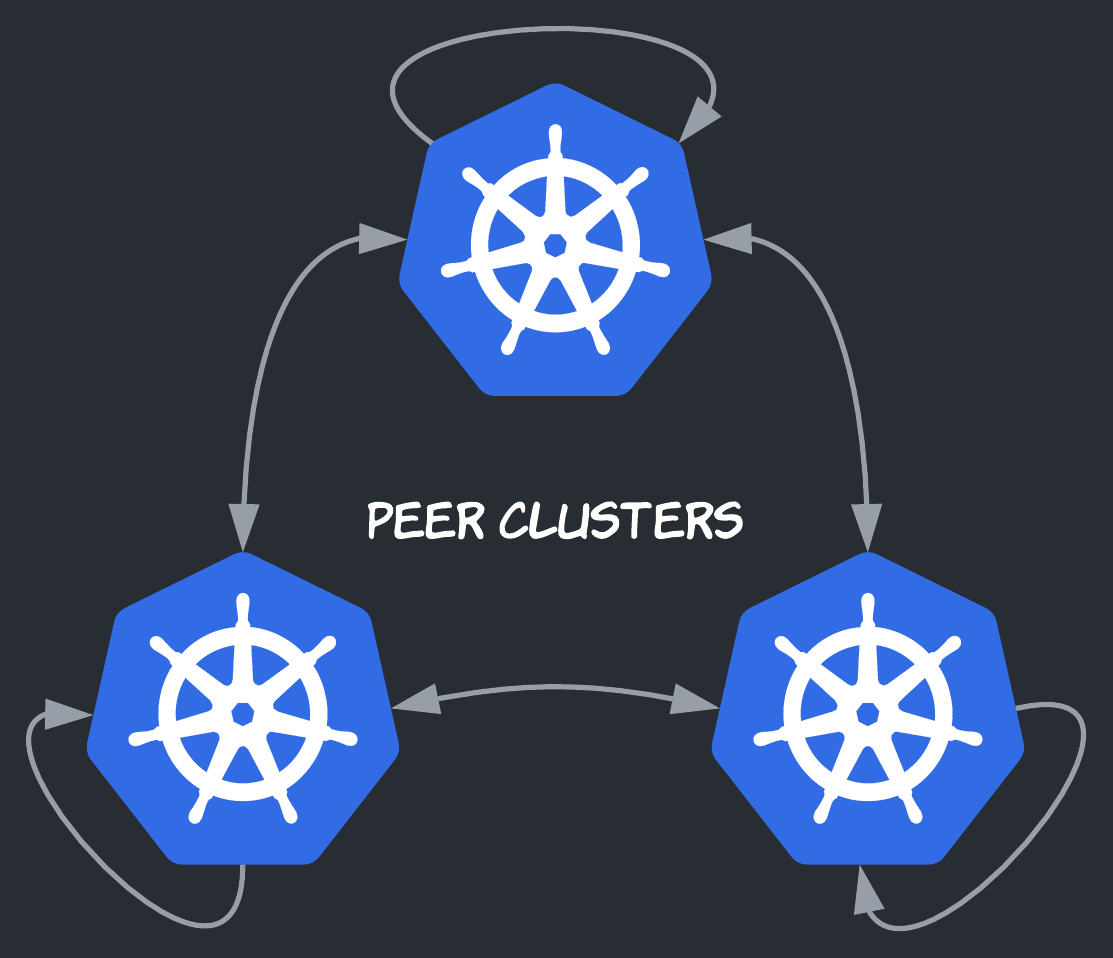}
    \caption{Admiralty decentralized federation}
    \label{fig:decentralized}
\end{figure}

Clusters and their connections form a directed graph. In particular:

\begin{itemize}
\item Saying that cluster A is a source of cluster B is equivalent to saying that cluster B is a target of cluster A.
\item Each cluster can be a source and/or a target of multiple clusters.
\item Connections can go both ways.
\item Clusters can target themselves.
\end{itemize}

This allows forming multiple federation topologies. Clusters can form a central control plane, where a single cluster is controlling several others (Figure~\ref{fig:central_plane}). This is common for other types of federation currently existing in Kubernetes ecosystem. It is also possible to form a burst type of federation, where workloads will optionally go to another cluster, most commonly into cloud. This is useful to offload some heavier types of workloads when the local cluster resources are not enough to serve the demand (Figure~\ref{fig:cloud_burst}). The most commonly used in Nautilus type of federation is decentralized, in which every cluster is on the same level with others, and users are creating connections that fit their needs in all possible directions (Figure~\ref{fig:decentralized}).


To run an offloaded workload in already federated namespace, a user should properly annotate the workload (Kubernetes pod, job or deployment), which will signal to Admiralty that it should consider other cluster for this workload. The Admiralty scheduler will then make a decision for placement and will go through the workflow described above to launch the workload.




The architecture of Admiralty makes it unique among other federation technologies in the way it works with Kubernetes. While others are adding additional layers on top of Kubernetes, Admiralty integrates with existing infrastructure and API. The result is the absence of additional client tools and abstractions, where the same API is used to send workloads between clusters and retrieve the results. \textbf{Admiralty is a native way for Kubernetes clusters to talk to each other.}

\section{Case Study: Wildfire Modeling at the Digital Continuum}
\begin{figure}
    \centering
    \includegraphics[width=0.47\textwidth]{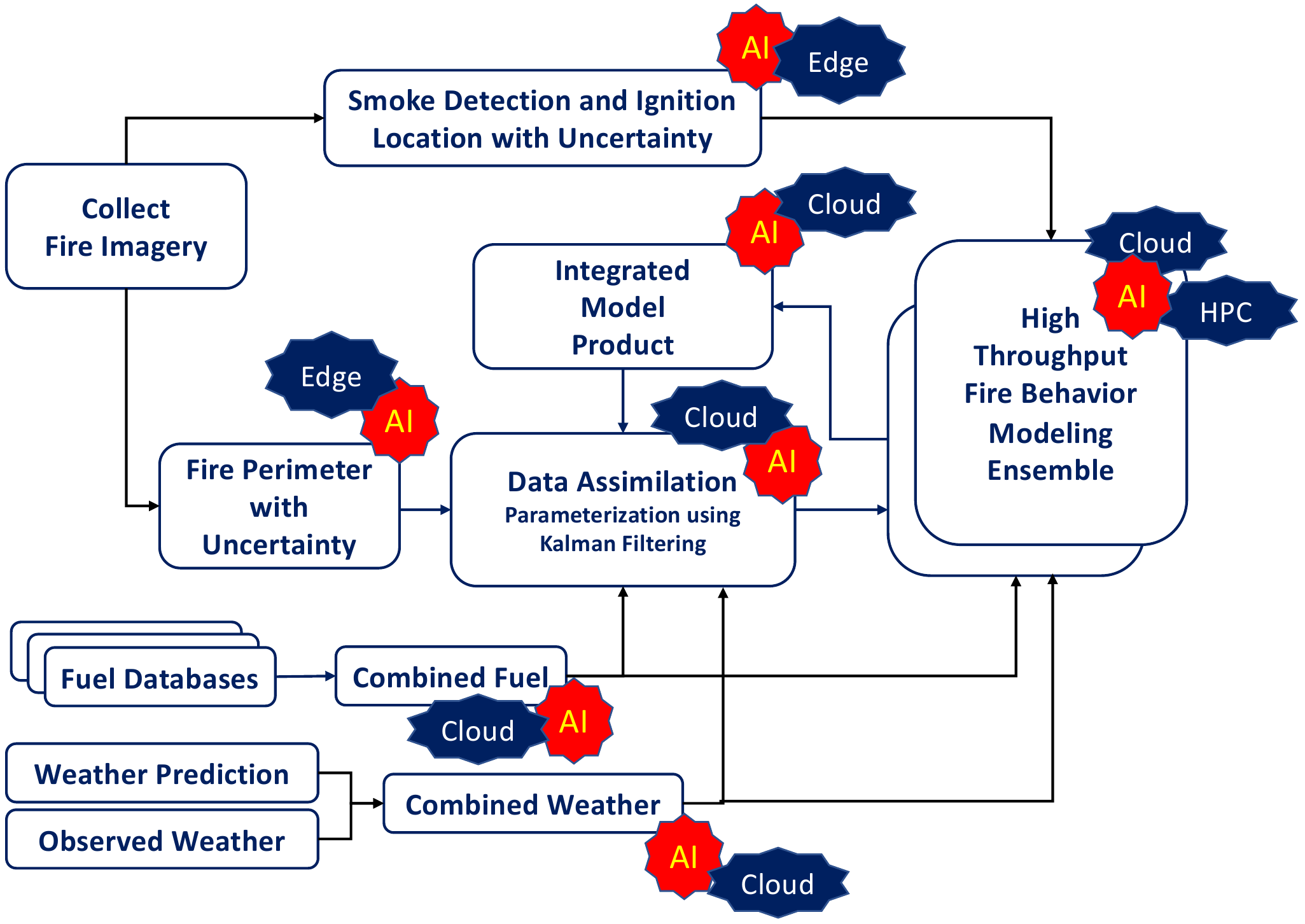}
    \caption{A conceptual set of steps for AI-driven fire modeling ensembles at the digital continuum from edge to cloud to HPC.}
    \label{fig:wifirewf}
    \vspace{-1ex}
\end{figure}

\begin{figure*}
    \centering
    \includegraphics[width=0.65\textwidth]{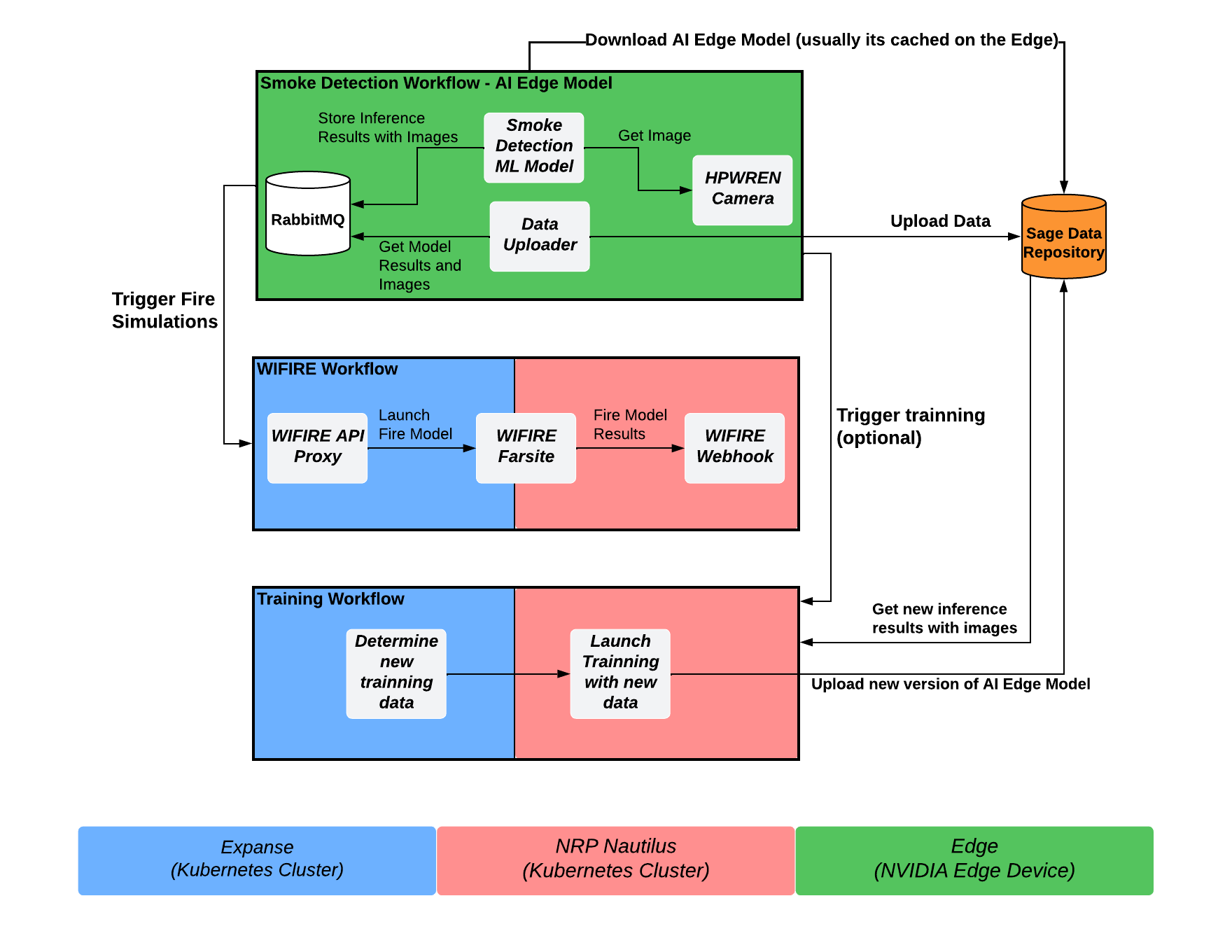}
    \caption{Composition of wildfire smoke detection and localization at the edge with fire modeling on the cloud and HPC.}
    \label{fig:workflowSmokeDetec}
\end{figure*}

We present a heterogeneous workflow that spans multiple compute environments from Edge to HPC to solve a complex wildland fire science problem.

Wildland fires have destroyed 4{\%} of California in 2020, the worst fire season in history. A common observation is that major destruction by fires is a worldwide trend,  and fire behavior is changing due to a combination of natural and anthropogenic factors, and frequency of catastophic fires is increasing \cite{firetrends}. Preventing the propagation of wildfires relies heavily on observations to be able to deploy resources and firefighters to combat the spread of wildfires. 
There is no doubt that having the right tools to combat both the initiation and propagation of wildfire is key to preventing events present an economic burden to our economy. This case study provides the next step to providing firefighters with the right tools to combat wildfires. The NSF-funded WIFIRE project \cite{wifire15} (wifire.ucsd.edu) took the first steps to tackle this problem, successfully creating an integrated system for wildfire monitoring, simulation, and response. Today, WIFIRE provides an end-to-end management cyberinfrastructure (CI) with integrated data collection from many real-time and archived data sources, knowledge management through AI, and modeling wildfire behavior using a plethora of community-developed wildfire modeling and simulation services at the digital continuum using many modes of computing.

Wildfire behavior modeling uses data involving field observations of fire, predicted and observed weather, and fuel data coming from various sources. Given a fire ignition point or an ongoing fire perimeter as a polygon, the forward spread of a fire from that point or polygon can be modeled using weather, vegetation and landscape data for that location. Although there are opportunities to use composable microservices and AI at different steps of such an integrated workflow as illustrated in Figure~\ref{fig:wifirewf}, in this case study, we focus on smoke detection and localization using mountaintop camera images and modeling the behavior of the fire based on the localized ignition point in a virtual and proof-of-concept scenario.

Automatic detection of a fire ignition or the perimeter of an ongoing fire is still a research topic that combines capturing imagery through remote sensing and analyzing the captured imagery through AI \cite{Govil_2020}. Any infrastructure built to enable these research efforts should provide support for data hosting, model training, model deployment and inference steps of the associated AI process. However, once reliable AI methods for this purpose are available, they have the potential to provide unprecedented edge intelligence capabilities in this domain. This potential made real-time smoke detection also one of the scientific case studies for the Sage AI on the edge project \cite{sage-wildfire}.

Sage \cite{SAGE} is an NSF-funded cyberinfrastructure that supports AI on the Edge with the primary aim to move AI to the edge. To this end, Sage deploys sensors with associated accelerators that can executed ready to schedule ML models, with an end goal to provide more than just a sensor value but inference-based insights based on the application's goal. The Sage software stack is a big component of this case study which further advances the scientific needs for wildfire science.

Figure~\ref{fig:workflowSmokeDetec} shows the architecture of the Smoke detection and wildfire modeling closed loop workflow. The rest of this section describes how this workflow was made possible in a composable system using a virtual deployment of Sage on Nautilus and WIFIRE's fire modeling services across the federated clusters.

\subsubsection{Edge}

The edge compute environment consists of the sensor data acquisition system (HPWREN cameras \cite{HPWREN}) and two Sage plugins that are part of the Sage software stack on a blade edge server. The first plugin is a camera plugin that is acquiring images at a higher frame rate than the normally deployed HPWREN camera and providing images to the other plugin. The smoke inference plugin (the second plugin) is a trained deep learning model that classifies acquired image from the previous plugin as either smoke or no smoke with a specific confidence level. The smoke plugin will then generate a message with basic information related to the inference done by the AI pre-trained model and send the message to the Sage Data Sensor Storage. The AI pre-trained model is obtained from the cloud through the Sage REST API which is fetching data from the S3 object storage denoted as Sage Object storage. Both the Sage Object Storage and Data Sensor Storage are part of Ceph which are shared between Expanse and Nautilus. 

The composable aspect of this system is highlighted in the Expanse and Nautilus sections of the architecture shown in Figure~\ref{fig:workflowSmokeDetec}. This is due to the nature of the demanding computing requirements to run an ensemble of fire simulations per each camera on either Expanse or Nautilus (cloud side).

\subsubsection{Expanse}
The first component is the training process which required a GPU node. The second component is the WIFIRE simulation workflow which only require a large memory CPU-based node. Expanse and Nautilus shared a common namespace called Enthalpy which is where the training and simulation jobs are being federated through Admiralty. Depending on the availability of either compute resource on each cluster, the workflow will be executed via the federation layer of the Composable System. 

During the training process a convolutional neural network model is trained with data from the Fire Ignition Library \cite{HPWRENImLib} which is an open source image library that consist of historical fires that were labeled smoke or no smoke 40 minutes before and after the time that a fire started and become visible from the HPWREN cameras. The library is also stored in the Sage Object Storage and accessible to both Expanse and Nautilus through either the REST API or Ceph. Furthermore, the training process is deployed on Enthalpy which can either run on Expanse or Nautilus. Due to the mature of the problem the time to solution needs to be minimized which is possible with the federation layer. In addition, the re-training step is triggered by the smoke detection message from the Edge with high confidence level and the training algorithm is executed as a batch process. This way, the newly acquired images (with high accuracy of smoke) from the Edge are used to continuously re-train the model. This allows us to minimize the need of human-labeled images and acquire more images.

In the course of the wildfire workflow, the smoke detection message from the Edge launches the fire computational model with the contents of the message as the input to the model to simulate the propagation of the fire. For this workflow, the compute needs are only CPU based.

\subsubsection{Nautilus}
Aside from the deep learning training and wildfire simulation workflow, Nautilus also host the Sage microservices such as the Sage REST API and the Lambda Trigger. The Lambda Trigger is a microservice that is responsible to launch a new ML model on newly acquired image data or to run the wildfire simulation workflow from the ML inference done on the Edge. The Sage REST API is a way to access the smoke detection deep learning model and training data for an outside services that is not part of Ceph. This is primarily the smoke detection plugin on the Edge that needs to have a trained ML model to perform the inference.

\section{Related Work}
Although the concept of a composable system is not necessarily new, this work presents, for the first time, the composition of an XSEDE (https://www.xsede.org/) supercomputer that has both a Kubernetes and Slurm partitions (Expanse) and a Kubernetes cluster (Nautilus). This new type of system will enable scientist to tackle their domain-specific problems with scientific workflows that can seamlessly produce closed loop workflows. Furthermore, scientists that have an XSEDE allocation will have access to a composable system and further push the boundaries of their scientific application.

There are a couple of APIs designed or systems that are using federation across multiple Kubernetes clusters and can potentially create a composable system. Consul(https://www.consul.io/) from Hashicorp federates multiple Kubernetes clusters and also between Kubernetes clusters and virtual machines.  
Openstack's API for federation across multiple Kubernetes cluster is called Magnum Federation API.  
The upbound.io federation tool crossplane.io is great for managing several clusters running in cloud providers (AWS, GCP, Azure, Alibaba) from another cluster to create multiple custom resources and is not useful for single-level connections between clusters in TNRP, and makes it hard to federate clusters with different ownerships. 

 Composable system described in this paper helps to build the distributed workflows across several clusters, requiring single-level connections between clusters. Although the use of  above-mentioned systems and APIs should be investigated for such use, at the time of the development of the paper, no such workflows were available.



\section{Conclusions and Future Work}

This paper describes a federation architecture to enable a composable infrastructure to ensure seamless integration of heterogeneous compute and storage resources using the Kubernetes ecosystem, ensuring interoperability, dynamic scalability, interactivity and performance monitoring of applications workflows. An application pattern involving AI data pipelines and processes for model training, inference and deployment, and combining the AI processes with big data and real-time simulation requirements were discussed. In alignment the eScience conference focus and audience, the paper is meant as an overview of a case study that highlights the design and use of composable systems. The contributions are significant from this implementation point of view as this is the first time such a heterogeneous workflow was implemented across multiple NSF resources. 

We believe the wildfire heterogeneous workflow is representative of many applications and paves the way for usage of composable systems in a generalizable fashion including: (\textit{i}) AI-driven simulations, such as the AI-driven exploration of time-dependent dynamics of molecular systems \cite{Casalino2020.11.19.390187}, to be deployed seamlessly; (\textit{ii}) big data analysis at scale, e.g., AI-integrated analysis of time-series data coming from wearables \cite{Smarr2020}; and (\textit{iii}) dynamic-data driven applications for urgent scenarios requiring large-memory and on-demand data parallelism at the same time, e.g., \cite{9041726}.

Performance analysis related to the case study workflow is left out of focus. Careful measurement and analysis of a step by step AI-integrated composable workflow can enable reduction of execution time and energy use significantly through dynamic resource configuration via Kubernetes. Grafana dashboards (e.g., Figure~\ref{fig:nu}) are used to continuously monitor resource usage

As part of the future directions, we will integrate the performance measurements collected from composed clusters through a dashboard that provides visualization and health monitoring of the infrastructure and services. We will provide user-facing tools for heterogeneous workflow development, deployment and scheduling process integration in a way that aligns with and supports a multi-cluster user allocation and authentication process. Moreover, we will focus on storage federation in networks of clusters, which is not possible in the current architecture. 

\section{Acknowledgments}

The authors would like to thank the WIFIRE and WorDS teams for their collaboration and support of the case study. Expanse , PRP, CHASE-CI and SAGE are supported through the NSF grants 1928224, 1541349, 1730158 and 1935984, respectively. The wildfire modeling case study was supported by NSF grants 1331615 and 2040676.

\bibliographystyle{IEEEtran}
\bibliography{2021IEEEEScience-ComposableSystems}

\end{document}